\newif\iflandscape
\newif\ifportrait
\typein[\lorp]%
{Typein "l" (for landscape, twocolumn) or "p" (for portrait, onecolumn)}
\if l\lorp \landscapetrue \else \portraittrue \fi
\newlength{\extralineskip}
\ifportrait
  \documentstyle[12pt]{article}
\fi
\iflandscape
  \documentstyle[twocolumn]{article}
\fi

\begin{document}
\begin{titlepage}
\begin{flushright}
          \begin{minipage}[t]{12em}
          \large UAB--FT--339\\
                 April 1994
          \end{minipage}
\end{flushright}

\vspace{\fill}

\vspace{\fill}

\begin{center}
\baselineskip=2.5em

{\LARGE  CONSTRAINTS ON  \\
NEUTRINO-NEUTRINO INTERACTIONS \\
FROM PRIMORDIAL NUCLEOSYNTHESIS}
\end{center}

\vspace{\fill}

\begin{center}
{\sc Eduard Mass\'o and Ramon Toldr\`a}\\

     Grup de F\'\i sica Te\`orica and Institut de F\'\i sica d'Altes Energies\\
     Universitat Aut\`onoma de Barcelona\\
     08193 Bellaterra, Barcelona, Spain
\end{center}

\vspace{\fill}

\begin{center}

\large ABSTRACT
\end{center}
\begin{center}
\begin{minipage}[t]{36em}
We use the constraints arising from primordial nucleosynthesis to bound
the strength $F$ of non-standard neutrino-neutrino interactions,
when the right-handed neutrinos participate in the interaction. We find
$F < 3 \times 10^{-3}\ G_F$, which is five orders of magnitude more
stringent than the limit obtained using LEP data. We also show that
secret interactions of neutrinos mediated by massless particles must
have a coupling $f$ less than $2 \times 10^{-5}$. This also
ameliorates previous limits in the literature.
\end{minipage}
\end{center}

\vspace{\fill}

\end{titlepage}

\clearpage


\addtolength{\baselineskip}{\extralineskip}
In the standard electroweak model the coupling of neutrinos to the
neutral Z-boson is determined by the Lagrangian
\begin{eqnarray} \label{sm1}
{\cal L} & = &  \frac{-g}{2 \cos \theta_w} \ J^\mu Z_\mu \ ,
\nonumber \\
J^\mu & = & \sum_{e,\mu,\tau} \bar \nu_L \gamma^\mu \nu_L   \ .
\end{eqnarray}
{}From these couplings, it follows that the standard interaction between
neutrinos at low energies is given by the effective Hamiltonian
\begin{equation} \label{sm3}
{\cal H} = \frac{G_F}{\sqrt 2} \ J^\mu J_\mu  \ ,
\end{equation}
where $G_F$ is the Fermi constant.

For obvious reasons, it is difficult to subject the $\nu-\nu$ effective
interaction to experimental scrutiny. Yet one would like
to test the standard interaction between neutrinos as much as possible. With
such a purpose one introduces non-standard interactions of the type
\cite{Bardin}
\begin{equation} \label{ns1} {\cal H}  =  F_V (\bar \nu_i \gamma^\mu
\nu_i)(\bar \nu_j \gamma_\mu \nu_j)
  + ...
\end{equation}
(a sum over $i,j=e,\mu,\tau$ is understood). In the last equation we have
only displayed a vector-vector type term. We have no reason to
limit ourselves to such a type of terms, and the dots in eq.(\ref{ns1})
refer to other possible tensorial forms as scalar, axial, etc...

{}From the phenomenological point of view, a test of the standard interactions
in eq.(\ref{sm3}) is performed by assuming that, in addition
to the standard Hamiltonian, one also has the new effective Hamiltonian
of eq.(\ref{ns1}). Then one uses experiments and observations that
allow to bound the strength $F_V$ of the non-standard interactions.

Upper bounds on $F_V$ have been calculated in ref.\cite{Bardin} by using some
$\pi$ and $K$ decays and neutrino reactions as well as in ref.\cite{Cable}
by analyzing $K \rightarrow \mu + 3\nu$. The neutrino data from SN1987A
have also been used to constrain non-standard interactions of
the neutrinos \cite{Manohar,Turner}.

The most stringent upper limit has been found recently in ref.\cite{Bilenky}
and it follows from the measure of the invisible
width of the $Z$-boson at LEP. A
non-standard four-neutrino coupling gives rise to the decay $Z \rightarrow \nu
\bar \nu \rightarrow \nu \bar \nu \nu \bar \nu$, which modifies the invisible
$Z$ width as predicted by the standard electroweak model. It is found that
\cite{Bilenky}
\begin{equation} \label{LEP}
F_V < 4 \times 10^2 \ G_F \ .
\end{equation}
Other types of non-standard $\nu-\nu$ interactions (axial, scalar, etc.)
are constrained by limits of the same order of magnitude that the one in
eq.(\ref{LEP}). The conclusion is that present laboratory and observational
limits still allow a non-standard interaction between neutrinos that is two
orders of magnitude stronger than the standard coupling given by the Fermi
constant.

In the present letter we will show that primordial nucleosynthesis constrains
non-standard interactions of neutrinos with bounds that are five orders of
magnitude more stringent than eq.(\ref{LEP}). For our bounds to hold, we
have to assume that the three neutrinos are Dirac fermions with masses
$m_\nu<<1$ MeV, and that the right-handed degrees of freedom participate in
the new interactions.

Primordial nucleosynthesis offers a limit on the degrees of freedom
contributing to the early universe expansion for $T \geq 1$ MeV. Three
right-handed (RH) neutrinos in equilibrium with their left-handed partners at
these temperatures are completely excluded \cite{Walker}. In fact, the three RH
neutrinos should have decoupled by the time the universe has temperatures of
about the QCD transition temperature $T \approx 200$ MeV. Indeed, at about
that time reactions heating the photons occur, and the relatively cold RH
neutrinos contribute much less to the expansion rate of the universe and do
not spoil the successful predictions of primordial nucleosynthesis.

The processes that can keep a RH neutrino $\nu_{Ri}$ in thermal equilibrium due
to the non-standard couplings are
\begin{equation} \label{reactions}
\bar \nu_{Li} + \nu_{Ri}  \longleftrightarrow  \bar \nu_{Rj} + \nu_{Lj}
\ ,
\end{equation}
where $j=e,\mu$ or $\tau$. The corresponding cross section is
\begin{equation} \label{sigma}
\sigma =  \frac{1}{4\pi} F^2_V s  \ .
\end{equation}
 The interaction rate (per unit time and per
$\nu_{Ri}$) is given by
\begin{equation} \label{rate}
\Gamma = n  \, <\sigma v>  \ .
\end{equation}
with the number density of $\bar \nu_{Li}$ neutrinos
\begin{equation} \label{n}
n =  \frac{3}{4} \frac{\xi(3)}{\pi^2} T^3   \ .
\end{equation}
We approximate $<\sigma v>$ by
\begin{equation} \label{sigma_FF}
<\sigma v> \ \approx  \ \frac{1}{2\pi} F_V^2 <s> \ ,
\end{equation}
and
\begin{equation} \label{s}
\sqrt{ <s> } \ \approx \ <E_\nu> \, =  \frac{ 7 \pi^4 T}{180 \xi(3)} \ .
\end{equation}
These approximations will lead to the correct order of magnitude for the
bound on $F_V$. Since our purpose is to show that one
can get a considerable improvement on the
constraints on non-standard interactions using primordial nucleosynthesis,
it is enough to approximate the average of the cross section as we have
done.

Enforcing that there is decoupling at a temperature $T$ amounts to impose
\begin{equation} \label{dec}
\Gamma  < H  \ ,
\end{equation}
with the Hubble parameter as a function of temperature given by
\begin{equation} \label{h}
H^2 = \frac{4\pi^3}{45} \frac{1}{M_P^2} g_* T^4 \ .
\end{equation}
Introducing $\Gamma$ from eq.(\ref{rate}) and working at $T = 200$ MeV,
with the corresponding relativistic degrees of freedom $g_*\simeq 50$,
we get from eq.(\ref{dec}) the limit
\begin{equation} \label{limit1}
F_V < 3 \times 10^{-3} G_F \ .
\end{equation}

Eq.(\ref{limit1}) is the first of our two main numerical results.
It says that non-standard
interactions between neutrinos have to be weaker than the standard weak
interactions by two or three orders of magnitude. Had we chosen another type
of interactions, scalar, V-A, etc., we would have obtained a limit of the same
order of magnitude than the one on $F_V$, eq.(\ref{limit1}).
We recall that our limit holds
when the RH neutrinos participate in the non-standard interactions.

The non-standard four-neutrino couplings could be the manifestation of
secret interactions of neutrinos, as introduced in ref.\cite{Turner}. These
new type of interactions would be beyond the standard model and not shared by
charged particles. They would be described by a Lagrangian of the form
\begin{equation} \label{ns2}
{\cal L} =   f_V (\bar \nu_i \gamma_\mu \nu_i) V^\mu + ...
\end{equation}
with $V$ a vector boson field  and the dots refer to other
possible terms, as scalar, axial couplings, etc...

The exchange of a massive $V$-boson gives rise to effective interactions at low
energy of the form shown in eq.(\ref{ns1}). For example, the bound of
ref.\cite{Bilenky} from LEP, eq.(\ref{LEP}),
can be interpreted as a bound on $f_V$
provided $m_V>>M_Z$. However, secret interactions of neutrinos with a
massless boson $V$ are not constrained by the analysis of ref.\cite{Bilenky}.

When the neutrino couples with strength $f_V$ to new bosons that are massless,
one may still place bounds as has been done in refs.
\cite{Manohar,Turner},
using information obtained from the observed $\nu$ pulse from SN1987A. The
most stringent limit arises when limiting the interaction of the neutrinos
in flight
to the earth with the neutrinos of the cosmic neutrino background. Assuming
all neutrinos are massless, it is found that \cite{Turner}
\begin{equation} \label{SN}
f_V < 5.6 \times 10^{-4}  \ .
\end{equation}
This limit is valid for masses $m_V<60$ eV.

Our limit from primordial nucleosynthesis,
eq.(\ref{limit1}) can be seen as a limit on
the secret interactions of neutrinos, provided $m_V >> 200$ MeV. When
neutrinos have secret couplings to a massless vector boson, we still can
constrain them using primordial nucleosynthesis,
 since RH neutrinos have now an interaction rate
\begin{equation} \label{gamma}
\Gamma   = n  \, <\sigma v> \ ,
\end{equation}
with
\begin{equation} \label{sigma_f}
<\sigma v> \ \approx \ \frac{1}{4\pi} \frac{f_V^4}{<s>}  \ .
\end{equation}

Introducing the relevant parameters we obtain
\begin{equation} \label{limit2} f_V < 2 \times 10^{-5} \ .
\end{equation}
This is our second main numerical result. Our bound is more stringent
than eq.(\ref{SN}) and it is valid for vector masses $m_V << 1$ MeV. In
fact, in this regime there are additional processes that may keep the RH
neutrinos in equilibrium, as $ \nu \bar \nu \leftrightarrow VV$. This
would make our bound in eq.(\ref{limit2}) slightly more stringent.

\vspace{2em}
This work has been partially supported by the CICYT
research project numbers AEN-93-0474 and AEN-93-0520. One of us (R.T.)
acknowledges a FPI grant from Ministerio de Educaci\'on y Ciencia
(Spain).

%
%

\vspace{4em}

\end{document}